## *FRACTIONAL DIFFERENTIAL and INTEGRAL OPERATIONS and CUMULATIVE PROCESSES*

F.Büyükkılıç[1], Z.Ok Bayrakdar[2] , D.Demirhan[3]

[1]fevzi.buyukkilic@ege.edu.tr [2] zahideok@gmail.com [3] dogan.demirhan@ege.edu.tr

Department of Physics, Faculty of Science, Ege University, Izmir, 35100 Bornova, Turkey

### *Abstract*

*In this study the general formula for differential and integral operations of fractional calculus via fractal operators by the method of cumulative diminution and cumulative growth is obtained . The under lying mechanism in the successof traditional fractional calculus for describing complex systems is uncovered. The connection between complex physics with fractional differentiation and integration operations is established.*

## *1.Introduction*

*Ordinary calculus and standard functions are inadequate to describe the complex systems. Fractional calculus is successfull in describing stochastic systems which have long-time memory and long-range interaction West et al. (2003); Barkai et al. (2000); Metzler, Klaftler (2000); Mandelbrot and Ness (1968); Luo and Afraimovich (2010); Beran (1994); Montroll and West (1978); Montroll and Shlesinger (1983), Marek-Crnjac et al. (2009); Klafter et al. (1990); Lindenberg and West (1990); Carpinteri and Mainardi (1991); Hilfer (2000); Stanislavsky (2004); West and Grigolini (1998); Sandev et al.(2014); Viñales et al. (2009); Fa (2006); Fa (2007); Camargo et al. (2009); Camargo et al. (2009); Camargo et al. (2009); Viñales and Despósito (2006); Viñales and Despósito (2007); Viñales et al. (2009); Scalas (2006); Klafter et al. (1995). In this work, the underlying mechanism in the basis of the fractional differintegral is put forward by the cumulative diminution and growth processes. It is demonstrated that in the real processes which evolve in a fractal space and discrete time with a non-Markovian memory, the mathematical descriptions are realized by a Fibonacci spirit simply by compound diminuation, compound growth, Buyukkilic and Demirhan (2008); Buyukkilic and Demirhan (2009); Kak (2004); Mandelbrot (1982); Stakhov (2005); Stakhov (2006); Stakhov (2006); Stakhov (2006). Within this approach, well known rate equation, the Brownian motion and the Langévin equation are handled involving fractal operator of derivation and Fibonacci method Bazzani et al. (2003); Doob (1942); Jumarie (2008);*

*Tarasov and Zaslavsky (2006); Tarasov (2012); Mazo (2002); Coffey et al. (2004). The general formulas of fractional differential and integral operations are obtained . Meanwhile , the physical mechanism underlying in the success of the fractional calculus for handling the complex systems has been uncovered mainly by compound diminution and growth approximations . The sections in this manuscript are organised as follows. In section 2 an introduction to fractal operator model of cumulative diminution and growth phenomena is outlined. In section 3 it is emphasized that cumulative processes can be treated as an eigenvalue problem . In section 4, the solution of the Langévin equation by compound diminution method is given. In section 5, investigation of Langévin equation in the framework of fractional calculus is handled. In section 6, in order to obtain the general formula for differential and integral operations of fractional calculus via fractal operators by the method of cumulative diminution and cumulative growth is stated. Section 7 is devoted to Cumulative Approach to the Integral Calculus. Section 8, a common formulation of derivative and integral calculations is given. And finally the results and discussion are pointed out.*

## 2.Introduction to Fractal Operator Model of Cumulative Diminution and Growth Process

*After the introduction of the analytical solutions of the cumulative diminution and growth processes, it is now convenient to introduce fractal operators for the processes to be handled in a general framework. In this framework , let us first consider the diminution process using operator technique. Let* $A_0$ *be the present value of the physical quantity at time* $t=0$. *For the time increments* $\Delta t, 2\Delta t, 3\Delta t, .., r\Delta t, \dots n\Delta t$ *, the corresponding compound values are* $A_1, A_2, A_3, \dots, A_r, \dots A_n$ *respectively. In this process, the quantity* $A_r$ *is obtainedfrom* $A_{r-1}$ *by subtracting from the* $A_{r-1}$ *the quantity that is deduced by the action of the operator* $\hat{B}$ *on the quantity* $A_{r-1}$. *Thus, for the* $n$ *th* $(n=0,1,\dots,)$*fold of the time interval* $\Delta t$

*for the* $0^{th}$ *step* $\quad A_0 = A_0$

*for the* $1^{st}$ *step* $\quad A_1 = A_0 - \hat{B}A_0 = \hat{C}A_0$

*for the* $2^{nd}$ *step* $\quad A_2 = A_1 - \hat{B}A_1 = \hat{C}^2A_0$

*for the* $r^{th}$ *step* $\quad A_r = A_{r-1} - \hat{B}A_{r-1} = \hat{C}^rA_0$

$\vdots \qquad\qquad \vdots$

*for the* $n^{th}$ *step* $$A_n = A_{n-1} - \hat{B}A_{n-1} = \hat{C}^n A_0 \qquad (1)$$

*can be written. Consequently, the relation between the future value* $A_n$ *and the present value* $A_0$ *is constructed by the fractal operator, West et al. (2003), Hilfer (2000), West and Grigolini (1998), West (1999) which is defined by:*

$$\hat{C}^n = (1 - \hat{B})^n . \qquad (2)$$

*To realize the effect(action) of the operator* $\hat{C}^n$ *on the quanity* $A_0$ *, the operator* $\hat{C}^n$ *is expanded into a Binomial series*

$$\hat{C}^n = \sum_{k=0}^{n} \binom{n}{k} \left(-\hat{B}\right)^k \qquad (3)$$

*where,* $\binom{n}{k}$ *are the Binomial coefficients. When the equation (3) is substituted into equation (1), then*

$$A_n = \sum_{k=0}^{n} \binom{n}{k} \left(-\hat{B}\right)^k A_0 \qquad (4)$$

*is obtained. If the Binomial coefficients* $\binom{n}{k}$ *in equation (4) are rewritten explicitly, after* $n$ *th step at time* $t = n\Delta t$*, the future value for the quantity* $A$ *will be as follows:*

$$A_n = \sum_{k=0}^{n} \frac{n(n-1)\ldots(n-k+1)}{k!} \left(-\hat{B}\right)^k A_0 \quad . \qquad (5)$$

*The inverse of the operator which is described in equation (2)*

$$\hat{C}^{-n} = \left(1 - \hat{B}\right)^{-n} \qquad (6)$$

*can be defined. By considering the fact that the cumulative diminution process is reversible, the inverse operator which is given in equation (6) is operated on to the both sides of equation (1) . Then*

$$A_0 = (1 - \hat{B})^{-n} A_n \qquad (7)$$

*is deduced. Consequently, from equation (7) , it is observed that the present value* $A_0$ *can be determined in terms of the future values. On the other hand by using a Binomial expansion, equation (7) can be written an alternative form :*

$$A_0 = \sum_{k=0}^{n} \binom{n+k-1}{k} \hat{B}^k A_n \,. \tag{8}$$

*When the properties of the Gamma function are used, the present value is determined as*

$$A_0 = \sum_{k=0}^{n} \frac{\Gamma(n+k)}{\Gamma(k+1)\Gamma(n)} \hat{B}^k A_n. \tag{9}$$

*Consequently, by fractal operator technique, the relation between the present value of the physical quantity $A_0$ and its future value $A$ is constructed .*

***3. Cumulative Processes as an Eigenvalue Problem***

*As an application of the diminution process which is mentioned above in section 2, let us consider the following eigenvalue equation :*

$$\hat{B} A_n(t) = \lambda \Delta t A_n(t). \tag{10}$$

*In this equation, eigenvalue $\lambda \Delta t$ is obtained as a result of action of the operator $\hat{B}$ on to the states $A_n(t)$. It is outlined in section 2 that after $n$ steps, initial value quantity $A_0$ is connected future value by the equation*

$$A_n(t) = \hat{C}^n A_0 \tag{11}$$

*where $\hat{C} = 1 - \lambda \Delta t$. At the n th step, where after $t = n\Delta t$ time amount elapses for the process to be realized, the quantity $A_n$ is obtained in terms of reduction rate $\lambda$ and time $t$ as :*

$$A_n(\lambda t) = \sum_{k=0}^{n} \frac{(-1)^k n(n-1) \dots (n-k+1)}{k!\, n^k} (\lambda t)^k A_0 \;. \tag{12}$$

*or one can write*

$$A_n(\lambda t) = \sum_{k=0}^{n} (-1)^k \frac{Q(n,k)}{k!} (\lambda t)^k A_0 \tag{13}$$

*where $Q(n,k)$ is defined by*

$$Q(n,k) = \prod_{i=1}^{k} \left(1 - \frac{k-i}{n}\right). \tag{14}$$

*For sufficiently large values of* $n$ , $Q(n,k) \to 1$. *When one of the properties of the Gamma function, namely* $k! = \Gamma(k+1)$*is used then,*

$$A(\lambda t) = \sum_{k=0}^{n} (-1)^k \frac{(\lambda t)^k}{\Gamma(k+1)} A_0 \tag{15}$$

*can be written. At this stage, the future value function* $A(\lambda t)$*can be rewritten in terms of Mittag- Leffler(M-L) function which is defined as Haubold and Mathai (2009); Mainardi and Gorenflo (2009); Kilbas and Saigo (1996); Oldham and Spanier (1974); Podlubny (1999); Kilbas et al (2006); Miller and Ross (1993)*

$$E_{\alpha,\beta}(z) = \sum_{k=0}^{\infty} \frac{z^k}{\Gamma(\alpha k + \beta)} \tag{16}$$

*then,*

$$A(\lambda t) = E_1(-\lambda t) A_0 \tag{17}$$

*is deduced.*

*It is observed that, the future value of a cumulative diminishing physical quantity which evolves in fractal space and discrete time, naturally leads to the Mittag-Leffler function. M-L function can be considered as a special* $\hat{C}^n$*operator. If* $\lambda t \ll 1$, *namely, product of the reduction rate and time is very small from 1, then the discrete evolution of the process is to be removed and lead to following equality*

$$A(\lambda t) = e^{-\lambda t} A_0. \tag{18}$$

*The result that is given in quation (18) is the same with the solution of the rate equation*

$$dA = -\lambda dt A. \tag{19}$$

*Then, in this work, the difference between the method which is used to describe the cumulative diminishing process and the standard approach is to be handled within the Fibonacci perception/ spirit. In other words, the relation of the steps entangled each other with respect to space and time, the evolution of the process as a non-Markovian manner is*

*going to be investigated . In our approach, process is evolved in fractal space-time as a diminution process. As it is known, the solution namely M-L function, which is deduced with fractional calculus, takes the place of the exponential function that is achieved from ordinary calculus. Morever, in this work it is observed that the future values and the initial values are bridged together in terms of M-L function.*

*The derivation of the present value from the future values is precious in regard to exhibit the irreversibility of time mathematically, so that the future is not realized in practice. With regard to this as a new application, let us obtain the present value from the future ones.*

*For the eigenvalue equation, in view of equation (8), the present value can be written in terms of the future values as*

$$A_0 = \sum_{k=0}^{n} \binom{n+k-1}{k} (\lambda\Delta t)^k A_n(\lambda\Delta t) \tag{20}$$

*Taking* $t = n\Delta t$ *, equation (20) can be rewritten in the form*

$$A_0 = \sum_{k=0}^{n} \frac{1}{n^k} \binom{n+k-1}{k} (\lambda t)^k A_n(\lambda t) \tag{21}$$

*or*

$$A_0 = \sum_{k=0}^{n} \frac{Q(n,k)}{k!} (\lambda t)^k A_n(\lambda t) \tag{22}$$

*where* $Q(n,k)$ *is defined as*

$$Q(n,k) = \prod_{i=1}^{k} \left(1 - \frac{k-i}{n}\right)$$

*For sufficiently large values of* $n$, $Q(n,k) \to 1$. *In this case equation (21) takes the following form*

$$A_0 = \sum_{k=0}^{n} \frac{(\lambda t)^k}{k!} A_n(\lambda t).$$

*Recalling the relation* $k! = \Gamma(k+1)$, *one gets*

$$A_0 = \sum_{k=0}^{\infty} \frac{(\lambda t)^k}{\Gamma(k+1)} A_n(\lambda t). \quad (23)$$

*In terms of M-L function, the relation between the initial value $A_0$ and the future value $A$ is obtained as*

$$A_0 = E_{1,1}(\lambda t) A_n(\lambda t) \quad (24)$$

*If $\lambda t \ll 1$, the initial value $A_0$ can be written, in terms of the future value of A by the following equation :*

$$A_0 = e^{\lambda t} A. \quad (25)$$

*It is observed that, it is possible to reach to the initial value, in a cumulative proceeding manner and starting by an increase from future value. Thus, this means that in the cumulative process, which is a turn around process and where the space is considered as fractal space , the event is reversible.*

*In fractional calculus, M-L function which takes the place of exponential function is confessed naturally in order to describe the fact that process proceeds in fractal space and discrete time. Thus, a clue is found for the mechanism which underlies in the success of the fractional calculus to describe the complex physical systems, namely, the concomitant physical quantity cumulative progress. Consequently, it is concluded that approaching to events by Fibonacci spirit leads us to more correct descriptions.*

## ***4.Solution of the Langévin Equation by Compound Diminution Method***

*Let us consider the random motion of Brownian particle in order to outline our cumulative diminution method. The Langévin equation of the Brownian particle which moves freely in a medium with an initial velocity $v_0$ , in the absence of an external force, can be written as a rate equation, Doob (1942), Mazo (2002), West (1999):*

$$\frac{dv(t)}{dt} + \lambda v(t) = 0 \quad (26)$$

*The velocity evolution of the Brownian particle can be handled with cumulative reduction method which is presented in section 2. At the time steps $t = 0, \Delta t, 2\Delta t, \dots, n\Delta t$, the values of the velocity can be modelled as*

$t = 0 \qquad v_0$

$t = \Delta t \qquad v_1 = v_0 - \lambda \Delta t v_0 = \hat{C} v_0$

$t = 2\Delta t \qquad v_2 = v_1 - \lambda \Delta t v_1 = \hat{C}^2 v_0$

$t = 3\Delta t \qquad v_3 = v_1 - \lambda \Delta t v_3 = \hat{C}^3 v_0$

$\vdots \qquad \vdots$

$$t = n\Delta t \qquad v_n = v_{n-1} - \lambda \Delta t v_{n-1} = \hat{C}^n v_0 \tag{27}$$

*where* $\hat{C} = 1 - \lambda \Delta t$ *. When the term* $\hat{C}^n$ *in equation (27) is expanded into a Binomial series*

$$v_n = \sum_{k=0}^{n} \binom{n}{k} (-\lambda \Delta t)^k v_0 \tag{28}$$

*is deduced. Substituting* $t = n\Delta t$*,in this expression leads to*

$$v_n(\lambda t) = \sum_{k=0}^{n} (-1)^k \frac{n(n-1)\ldots(n-k+1)}{k!\ n^k} (\lambda \Delta t)^k v_0 \tag{29}$$

*or*

$$v_n(\lambda t) = \sum_{k=0}^{n} (-1)^k \frac{Q(n,k)}{k!} (\lambda t)^k v_0 \tag{30}$$

*is obtained, where* $Q(n,k)$ *is defined by*

$$Q(n,k) = \prod_{i=1}^{k} \left(1 - \frac{k-i}{n}\right) \tag{31}$$

*For sufficiently large values of* $n$ , $Q(n,k) \to 1$*. Thereafter,*

$$v(\lambda t) = \sum_{k=0}^{\infty} (-1)^k \frac{(\lambda t)^k}{\Gamma(k+1)} v_0 \tag{32}$$

*can be written, where* $k! = \Gamma(k+1)$*.*

*In order to take into account the viscosity of the medium, instead of* $k$ *,* $k\alpha$ *must be substituted in equation (32) which gives us:*

$$v(\lambda t) = \sum_{k=0}^{\infty} (-1)^k \frac{(\lambda t)^{k\alpha}}{\Gamma(k\alpha+1)} v_0. \tag{33}$$

*It is well known that the property of the medium corresponds to the subdiffusion effect for* $0 < \alpha \leq 1$ *, superdiffusion effect for* $1 < \alpha < 2$ *and ballistic diffusion effect for* $\alpha = 2$ *West et al. (2003). In this investigation, taking into account the superdiffusion phenomenon ,* $0 < \alpha \leq 1$*,* $\alpha = 1$ *can be choosen. In this case, the evolution of the process in time is identical with equation (32).*

*At the beginning of the process, for* $\lambda t < 1$ *, in view of definition of M-L function given by (16), for* $t \to 0$ *the velocity is defined as*

$$v(\lambda t) = v_o E_1(-\lambda t) \to v_0 e^{-(\lambda t)^{\alpha}} \qquad (34a)$$

*and behave as streched exponential. In the proceeding steps of the process, at later times, namely when* $\lambda t$ *grows,i.e. for* $t \to \infty$ *the velocity evolves to*

$$v(\lambda t) = v_0 E_1(-\lambda t) \to v_\infty (\lambda t)^{-\alpha} \qquad (34b)$$

*inverse power law. As one can see M-L function extrapolates smoothly between the two asymptotic forms, West et al.(2003).*

## *5. Investigation of Langévin Equation in the Framework of Fractional Calculus*

*Let us now consider the evolution of the Brownian particle in the framework of fractional calculus which is presented above in section 4. Equation (26), can be written as an integral equation Hilfer (2000), Tarasov and Zaslavsky (2006), Nonnenmacher and Metzler (1995); Mathai et al. (2010); Glöcke and Nonnenmacher (1993)*

$$v(t) - v(0) = -\lambda^{\alpha} D_t^{-\alpha} v(t). \tag{35}$$

*In order to solve equation (35), Laplace transform method can be used. Thus, Laplace transform of equation (35) can be written as*

$$\tilde{v}(s) - \frac{v(0)}{s} = -\lambda^{\alpha} \mathcal{L}\{D_t^{-\alpha}\, ; s\}. \tag{36}$$

*The Laplace transform of fractional integral equation is given by*

$$\mathcal{L}\{D_t^{-\alpha}\,; s\} = \frac{\tilde{v}(s)}{s^{\alpha}} \tag{37}$$

*Tarasov and Zaslavsky (2006); Tarasov (2012); Mazo (2002); Coffey et al. (2004). Consequently, when the expression by (37) is substituted in equation (36), the following equation*

$$\tilde{v}(s) = v(0)\,\frac{(s)^{-1}}{1+\left(\frac{s}{\lambda}\right)^{-\alpha}} \tag{38}$$

*can be written West (1999); Haubold and Mathai (2009); Mainardi and Gorenflo (2009); Kilbas and Saigo (1996); Oldham and Spanier (1974); Podlubny (1999); Kilbas et al (2006); Miller and Ross (1993). Applying the Laplace transform operation to both sides of equation (38), time depended solution is found as*

$$v(t) = \mathcal{L}^{-1}\left\{v(0)\,\frac{(s)^{-1}}{1+\left(\frac{s}{\lambda}\right)^{-\alpha}}\,; t\right\}. \tag{39}$$

*As special cases,*

$$for \quad \alpha = 0 \qquad\qquad v(t) = v(0) \tag{39a}$$

$$for \quad \alpha = 1 \qquad\qquad v(t) = v(0)e^{-\lambda t} \tag{39b}$$

*can be written. For the explicit solution of (39), the denominator is expanded into MacLaurin series which then yields:*

$$v(t) = \mathcal{L}^{-1}\left\{v(0)\sum_{k=0}^{\infty}(-1)^k \lambda^{\alpha k} s^{-\alpha k-1}\right\}. \tag{40}$$

*Making use of Laplace transform,*

$$\mathcal{L}\{t^{\rho}\} = \Gamma(\rho+1)s^{-(\rho+1)} \quad , \qquad \rho > -1 \tag{41}$$

*the explicit solution of equation (40) is deduced as*

$$v(t) = v(0)\sum_{k=0}^{\infty}(-1)^k \frac{(\lambda t)^{\alpha k}}{\Gamma(\alpha k+1)}\,. \tag{42}$$

*In view of the definition of M-L function in (16), equation (42) can be written in the form*

$$v(t) = v(0)E_\alpha[(-\lambda t)^\alpha]\,. \qquad (43)$$

***6.Cumulative Approach to the Differential Calculation***

*In a cumulative diminution process which is mentioned above in section 2, let us choose the operator* $\hat{B}$ *as a differential operator. With the purpose of improving this approach to investigation of physical events with Fibonacci spirit where the basis of success of fractional calculus in describing these systems underlies, let us consider the differintegral operation with compound approach. In compound diminution process given in section 3, let us choose* $\hat{B}$ *in the form*

$$\hat{B} = e^{-\Delta t\hat{D}} \qquad (44)$$

*as a shift operator West et al. (2003), Hilfer (2000), West and Grigolini (1998). Thus, the effect of the operator* $\hat{B}$ *on the* $A(t)$ *is defined as*

$$\hat{B}A(t) = A(t-\Delta t). \qquad (45)$$

*In other words, operator* $\hat{B}$*, carries* $A(t)$ *to* $A(t-\Delta t)$*. Here, at the* $\Delta t, 2\Delta t, \dots, n\Delta t$ *time increments, the evolution of* $A_0(t)$ *is defined in this way*

*for the* $0^{th}$*step;* $\qquad A_0(t) = A_0(t)$

*for the* $1^{th}$*step;* $\quad \Delta t \qquad A_1(t) = A_0(t) - e^{-\Delta t\hat{D}}A_0(t) = \hat{C}A_0(t)$

*for the* $2^{nd}$ *step;* $\quad 2\Delta t \qquad A_2(t) = A_1(t) - e^{-\Delta t\hat{D}}A_1(t) = \hat{C}^2A_0(t)$ *(46)*

$\vdots \qquad\qquad \vdots$

*for the* $n^{th}$*step;*$n\Delta t \qquad A_n(t) = A_{n-1}(t) - e^{-\Delta t\hat{D}}A_{n-1}(t) = \hat{C}^nA_0(t).$

*Consequently, the relation between the present value* $A_0(t)$*and the future value* $A_n(t)$*is obtained.Thereby, this relation must be constructed with an operator* $\hat{C}$ *which is defines by*

$$\hat{C} = (1 - e^{-\Delta t\hat{D}}). \qquad (47)$$

*By making use of above definition, for $\Delta t \to 0$, a derivative operator $\widehat{D}$*

$$\lim_{\Delta t \to 0} \frac{\hat{C}}{\Delta t} = \lim_{\Delta t \to 0} \frac{1 - e^{-\Delta t \widehat{D}}}{\Delta t} = \widehat{D} \tag{48}$$

*can be written. To perform the derivative at $n$. thorder, operator $\hat{C}^n$ is expanded into a Binomial series:*

$$\hat{C}^n = \sum_{k=0}^{n} \binom{n}{k} \left(-e^{-\Delta t \widehat{D}}\right)^k . \tag{49}$$

*When equation (49)is substituted into equation (46) one gets*

$$A_n(t) = \sum_{k=0}^{n} \binom{n}{k} \left(-e^{-\Delta t \widehat{D}}\right)^k A_0(t). \tag{50}$$

*In view of the property of the shift operator of $\widehat{B}$ which is defined in equation (44), as the result of $k$ times operation of the operator , the following equality holds:*

$$(e^{-\Delta t \widehat{D}})^k A_0(t) = A_0(t - k\Delta t) . \tag{51}$$

*In this case, equation (50) can be written in the following form*

$$A_n(t) = \sum_{k=0}^{n} (-1)^k \binom{n}{k} A_0(t - k\Delta t). \tag{52}$$

*For $n = 1$ , from equation (52)*

$$A_1(t) = A_0(t) - A_0(t - \Delta t) \tag{53}$$

*can be written.*

*To perform differentiation calculus, equation (53) is divided by $\Delta t$ and the limit $\Delta t \to 0$ is taken into account. Then in this case one can write :*

$$\lim_{\Delta t \to 0} \frac{A_1(t)}{\Delta t} = \lim_{\Delta t \to 0} \frac{A_0(t) - A_0(t - \Delta t)}{\Delta t} = \lim_{\Delta t \to 0} \frac{\hat{C}}{\Delta t} A_0(t) . \tag{54}$$

*From the above equation*

$$\lim_{\Delta t \to 0} \frac{A_1(t)}{\Delta t} = \lim_{\Delta t \to 0} \frac{\hat{C}}{\Delta t} A_0(t) \tag{55}$$

*can be written down. By considering the definition of diferentiation given by (48), without indices,*

$$\frac{dA(t)}{dt} = \widehat{D}A(t) \tag{56}$$

*is obtained.*

*This definition of derivative can be generalized to higher order derivatives. For this purpose, by taking into account the Binomial expansion of* $\widehat{C}^n$

$$\frac{d^nA}{dt^n} = \frac{A_n}{\lim\limits_{\Delta t \to 0} \Delta t^n} = \lim_{\Delta t \to 0} \frac{1}{\Delta t^n} \sum_{k=0}^{n} (-1)^k \binom{n}{k} \left(e^{-\Delta t \widehat{D}}\right)^k A_0(t). \tag{57}$$

*standard series for the derivative is attained.*

*If the function* $A(t)$ *has* $n$ *th derivative,* $\Delta t$ *takes any value while achieving the limit in the derivation of derivative.*

*In order to write the definition of both the derivative and the integral together for the same function* $A(t)$*, for* $\Delta t \to 0$ *, the values are limited. Thus, let the function be defined in the interval* $(0, t)$ *and define* $\Delta t$ *as:*

$$\Delta t \equiv \frac{t - 0}{N} \ . \tag{58}$$

*For* $\Delta t \to 0$ *,* $N \to \infty$*. In this regard, equation (57) can be rewritten as*

$$\frac{d^nA}{dt^n} = \lim_{N \to \infty} \left(\frac{t}{N}\right)^{-n} \sum_{k=0}^{N-1} (-1)^k \binom{n}{k} A\left(t - k\frac{t}{N}\right) \tag{59}$$

*where, one of the properties of Binomial coefficients is used:*

$$\binom{n}{k} = 0, \qquad k > n\,. \tag{60}$$

## ***7.Cumulative Approach to the Integral Calculus***

*Now, let us derive the standard series expansion for the integral. In equation (7), if* $-n$ *is substituted instead of* $n$*, then from the cumulative diminution equation, present value* $A_0(t)$*can be written as*

$$A_0(t) = \left(1 - \hat{B}\right)^{-n} A_n(t) \tag{61}$$

*or*

$$A_0(t) = \sum_{k=0}^{N-1} \binom{n+k-1}{k} \hat{B}^k A_n(t). \tag{62}$$

*where, the propety of* $\binom{-n}{k} = (-1)^k \binom{n+k-1}{k}$ *is used.*

*Let us use a differential shift operator* $\hat{B} = e^{-\Delta t\hat{D}}$*in equation (62) which leads to:*

$$A_0(t) = \sum_{k=0}^{N-1} \binom{n+k-1}{k} (e^{-\Delta t\hat{D}})^k A_n(t). \tag{63}$$

*can be written. If this equation is multiply by* $(\Delta t)^n$*,then*

$$\frac{d^{-n}A(t)}{dt^{-n}} = \frac{A_0(t)}{(\Delta t)^{-n}} = \Delta t^n \sum_{k=0}^{N-1} \binom{k+n-1}{k} A_n(t - k\Delta t) \tag{64}$$

*is obtained. For* $\Delta t \to \frac{t}{N}$ *, n-fold integral of* $A(t)$ *is*

$$\frac{d^{-n}A(t)}{dt^{-n}} = \lim_{N\to\infty} \left(\frac{t}{N}\right)^n \sum_{k=0}^{N-1} \binom{k+n-1}{k} A\left(t - k\frac{t}{N}\right) . \tag{65}$$

*By writing the Binomial expansion explicitly and using theproperty of Gamma function*

$$(-1)^k \binom{n}{k} = \frac{\Gamma(k-n)}{\Gamma(k+1)\Gamma(-n)}$$

*the expression for derivative given in equation (57) is written as a series in the following form:*

$$\frac{d^nA(t)}{dt^n} = \lim_{N\to\infty} \left(\frac{t}{N}\right)^{-n} \sum_{k=0}^{N-1} \frac{\Gamma(k-n)}{\Gamma(k+1)\Gamma(-n)} A\left(t - k\frac{t}{N}\right). \tag{66}$$

*n th order integral when expressed as a series expansion is*

$$\frac{d^{-n}A(t)}{dt^{-n}} = \lim_{N\to\infty} \left(\frac{t}{N}\right)^n \sum_{k=0}^{N-1} \frac{\Gamma(k+n)}{\Gamma(k+1)\Gamma(n)} A\left(t - k\frac{t}{N}\right). \tag{67}$$

## *8. A Common Formulation of Derivative and Integral Calculus*

*$n$.th order derivative and $n$. th order integral expressions can be written down in a single expression. If equation (60) and (61) are cosidered together then $q^{th}$ order of differentiation or integration of the function A(t) reads:*

$$\frac{d^{\pm q}A(t)}{dt^{\pm q}} = \lim_{N\to\infty}\left(\frac{t}{N}\right)^{\mp q}\sum_{k=0}^{N-1}\frac{\Gamma(k\mp q)}{\Gamma(k+1)\Gamma(\mp q)}A\left(t-k\frac{t}{N}\right). \qquad (68)$$

*Here the sign ," +" is used for the derivative and the " -" sign is used for the integration operations. This equation is called differintegral and manifested by Oldham and Spanier (1974).*

## *9.Result and Discussions*

*Fractional mathematics is succesful in describing complex systems which evolve in fractal space and discrete time with a non –Markovian manner. However, the basic key formula of fractional calculus ; differintegral , remain within the domain of definition of mathematicians. Due to the importance of the fractional calculus, we attempted to the understanding of the underlying physical mechanism behind differintegral of fractional calculus.*

*In this study any particular phenomenon is not described in full detail. Fractal relaxation equation, have been used to model the evolution of stochastic phenomena with long-time memory ,that is , phenomena with correlations which decay with Mittag-Leffler and inverse power law types rather than exponentialy in time. Within this aim, fractal operator model of cumulative diminution and growth processes are introduced. In this framework,using fractal operators , the present and future values of a quantity are related. The general formulas of fractional calculus, differintegral which are given by several mathematicians are proved, by means of fractal operators of differentiation which are treated within the cumulative process of growth and diminution. We came to the conclusion that the underlying mechanismin the differintegral is cumulative growth and diminution process of Fibonacci appproach.*